\documentclass[modern]{aastex63}

\graphicspath{{./}}

\usepackage{amsmath}
\usepackage{float}

\begin{document}

\title{Constraints on the Abundance of PBHs from X-ray Quasar Microlensing Observations: Substellar to Planetary Mass Range}

\author{A. ESTEBAN-GUTI\'ERREZ} 
\affiliation{Instituto de Astrof\'{\i}sica de Canarias, V\'{\i}a L\'actea S/N, 38205 La Laguna, Tenerife, Spain}
\affiliation{Departamento de Astrof\'{\i}sica, Universidad de la Laguna, 38206 La Laguna, Tenerife, Spain}
\author{E. MEDIAVILLA}
\affiliation{Instituto de Astrof\'{\i}sica de Canarias, V\'{\i}a L\'actea S/N, 38205 La Laguna, Tenerife, Spain}
\affiliation{Departamento de Astrof\'{\i}sica, Universidad de la Laguna, 38206 La Laguna, Tenerife, Spain}
\author{J. JIM\'ENEZ-VICENTE}
\affiliation{Departamento de F\'{\i}sica Te\'orica y del Cosmos, Universidad de Granada, Campus de Fuentenueva, 18071 Granada, Spain}
\affiliation{Instituto Carlos I de F\'{\i}sica Te\'orica y Computacional, Universidad de Granada, 18071 Granada, Spain}
\author{J. A. MU\~NOZ}
\affiliation{Departamento de Astronom\'{\i}a y Astrof\'{\i}sica, Universidad de Valencia, 46100 Burjassot, Valencia, Spain.}
\affiliation{Observatorio Astron\'omico, Universidad de Valencia, 46980 Paterna, Valencia, Spain}

\begin{abstract}

 We use X-ray observations of quasar microlensing (sensitive to smaller compact objects than in the optical) to study the possible presence of a population of low mass black holes (from $\sim$ $10^{-3}M_{\odot}$ to $10^{-1}M_{\odot}$) in lens galaxies. We compare these observations with microlensing magnification simulations of a mixed population of stars and black holes (BHs) plus a smooth matter component. We estimate the individual mass fractions of both, stars and BHs, for three different BH masses in the range of substellar to planetary masses. Our Bayesian analysis indicates that the contribution of BHs is negligible in the substellar mass range but that a population of BHs of planetary mass (M $\lesssim$ $10^{-3}M_{\odot}$) could pass unnoticed to X-ray microlensing. 
 We provide new upper limits to the contribution of BHs to the fraction of dark matter based on both, the quasar microlensing data in the X-ray band, and our previous estimates in the optical of intermediate-mass BHs with an additional upper limit at $M=3M_{\odot}$.

\end{abstract}

\keywords{(Primordial Black Holes --- gravitational lensing: micro --- X-rays)} 

\section{Introduction \label{intro}}

    Since the first evidences of dark matter in galaxies and clusters of galaxies in the last century, the astrophysical community is still searching for a plausible candidate. Several possibilities have been proposed from different disciplines including elementary particles (Feng et al. 2010), new types of interacting dark matter (e.g. Salucci et al. 2020) or faint compact objects in the halos of galaxies (Alcock et al. 2000), but none of them have yet provided any strong evidence. 
    
    The discovery by the LIGO/Virgo gravitational wave (GW) collaborations of mergers of BHs in the $10-50M_{\odot}$ mass range (Abbott et al. 2019a, 2019b, 2021a, 2021b, The Ligo Scientific Collaboration et al. 2021a, 2021b) reopened the interest on primordial black holes (PBHs), theoretically predicted to be formed in the radiation-dominated era (Hawking 1971, Carr 1975) and postulated as a suitable dark matter candidate (Sasaki et al. 2018, Clesse \& Garc{\'\i}a-Bellido 2018). Nonetheless, those stellar-mass PBHs are not the only acceptable possibility, but a larger mass range ($\sim 10^{-12}-10^{3} M_{\odot}$) should be considered according to models of PBH formation (Carr \& Kühnel 2020, Carr et al. 2021b), with a particular focus on the promising mass window from 1$M_{\odot}$ down to $10^{-6}M_{\odot}$ (Carr et al. 2021a). 
    
    Gravitational microlensing, and specifically microlensing of lensed quasars, are ideal astrophysical phenomena to analyze the abundance of compact objects in lens galaxies (Chang \& Refsdal 1979, Schneider et al. 2006, Jim{\'e}nez-Vicente et al. 2015a). During recent years, the fraction of PBHs in the 1-100$M_{\odot}$ mass range and its possible contribution to dark matter have been strongly constrained by galactic microlensing (Alcock et al. 2001, Tisserand et al. 2007,  Griest et al. 2014, Zumalac\'arregui \& Seljak 2018, Wyrzykowski \& Mandel 2020, Verma \& Rentala 2022, Blaineau et al. 2022), quasar microlensing (Hawkins 2011, Mediavilla et al. 2017, Hawkins 2020, Esteban-Guti\'errez  et al. 2020, Esteban-Guti\'errez  et al. 2022a, Esteban-Guti\'errez  et al. 2022b, Hawkins 2022), GWs (Kavanagh et al. 2018, Vaskonen \& Veerm{\"a}e 2020) and galactic radio/X-ray emission (Inoue \& Kusenko 2017, Manshanden et al. 2019) studies. The majority of those works have estimated low mass fractions for PBHs, which could therefore constitute only a small fraction of the total dark matter content (see, nevertheless, Hawkins 2011, 2020, 2022).
    
    On the other hand, the low-mass range ($M \ll 1M_{\odot}$) has also been explored using galactic microlensing (e.g. OGLE\footnote{The Optical Gravitational Lensing Experiment, \url{http://ogle.astrouw.edu.pl}}, EROS\footnote{Exp\'erience pour la Recherche d'Objets Sombres, \url{http://eros.in2p3.fr}}, HSC\footnote{Hyper Suprime-Cam, \url{https://www.naoj.org}}), GWs (e.g. LIGO\footnote{Laser Interferometer Gravitational-Wave Observatory, \url{https://www.ligo.caltech.edu}}, Virgo\footnote{The Virgo interferometer, \url{https://www.virgo-gw.eu}}) or pulsars (e.g. NANOGrav\footnote{North American Nanohertz Observatory for Gravitational Waves, \url{https://nanograv.org}}). While the majority of these works found very low bounds to the PBH abundances in the substellar or subsolar mass regime (Tisserand et al. 2007, Oguri et al. 2018, Abbott et al. 2018, Niikura et al. 2019, Smyth et al. 2020, Chen et al. 2020, Chen \& Huang 2020, Abbott et al. 2022, The Ligo Scientific Collaboration et al. 2022), there are others that claim the possibility that PBHs in the $\sim 10^{-12}-100M_{\odot}$ mass range (Dror et al. 2019) or in the asteroid to planetary mass range (Miller et al. 2021, 2022, Dom{\`e}nech \& Pi 2022) could account for an important fraction of dark matter.
    
     If we want to explore small mass PBHs using quasar microlensing of lensed galaxies, we need to use very compact sources, with sizes smaller than the Einstein radius of the considered microlens masses. This can be achieved by using observations in the X-ray band, which could be sensitive to the effect of the smallest microlenses (e.g. Pooley et al. 2007, Jim{\'e}nez-Vicente et al. 2015a, 2015b). Thus, this work aims to add constraints to the dark matter fraction in form of compact objects in the substellar to planetary mass range from X-ray observations of quasar microlensing.
    
    The article is organized as follows. Section \S2 describes the data used for this analysis and the details of the specific parameters used in the microlensing simulations. The statistical tools applied to compute the probabilities are also defined in this section. In Section \S3, the main results found for the PBH abundances in the planetary to substellar mass range are presented. In Section \S4 we show and discuss our estimates of the contribution of PBHs to the fraction of dark matter in the context of previous works. Finally, the conclusions are outlined in Section \S5. \\

\section{X-ray Data and Methodology \label{data}}
    We use the X-ray differential microlensing magnifications collected by Jim{\'e}nez-Vicente et al. (2015b, see their Table 1) from the fluxes reported by Schechter et al. (2014) extracted from quasar observations by Pooley et al. (2007) and Blackburne et al. (2011). The macro model magnifications ($\mu_{i}$) provided by Schechter et al. (2014) are used as an unmicrolensed baseline. Therefore, the microlensing magnification between an image, $i$, and the reference image, $j$, of each system is given by:
    \begin{equation}
        \Delta m_{ij} =m_{i} - m_{j} - (\mu_{i} - \mu_{j})= (\Delta m_{i} - \Delta m_{j}),
    \end{equation} 
    
    Our selection consists then of a total of 30 quasar image pairs seen through 10 lens galaxies. To compare with these observations, we generate microlensing magnification maps produced by a mixture of stars and BHs, in surface mass density fractions $\alpha_{stars}=\kappa_{stars}/\kappa$ and $\alpha_{BH}=\kappa_{BH}/\kappa$, respectively, as we did in Esteban-Guti{\'e}rrez et al. (2022b). The mass of the stars is fixed to 0.2$M_{\odot}$, which is a representative value of the mean mass of the old stellar population in lens galaxies (see Poindexter \& Kochanek 2010, Jim{\'e}nez-Vicente \& Mediavilla 2019). The lowest mass for the BHs that we are able to probe is limited by the typical source size for the X-ray source of $\sim 1$ lt-day (see Jim{\'e}nez-Vicente et al. 2015b and references therein).
    Thus, the lower bound for the BH mass is taken to be the mass with an Einstein radius of size 1 lt-day\footnote{For a typical gravitational lens system with the lens at z=0.5 and the source at z=2.}, that is, $M_{BH} = 0.0024M_{\odot} (\sim 2.49M_{J})$. On the other side, the upper bound for the BH mass is that of the stars, as we are not able to distinguish BHs from stars above this limit. We therefore take as upper limit the Hydrogen Burning Limit (HBL) of $M_{BH} \sim 0.08M_{\odot}$. This way, we explore a BH mass range between the lowest suitable mass and the HBL in a logarithmic grid of masses as $M_{BH}/M_{\odot} = {0.0024, 0.013, 0.082}$, and perform three sets of simulations to generate the corresponding magnification maps.
    
    For the surface mass density fraction of the stars we use a linear grid of 7 values in $\alpha_{star} = \{0,0.05,0.1,0.2,0.3,0.4,0.8\}$. For the fraction of mass in BHs we take a grid of 7 values logarithmically distributed from $\alpha_{BH}=0.02$ to $\alpha_{BH}=1$, in addition to the the $\alpha_{BH}=0$ contribution, resulting in $\alpha_{BH} = \{0,0.02,0.044, 0.096, 0.21, 0.46, 1\}$. The additional contribution to the projected mass to complete the macro convergence, $\kappa$, is in form of a smooth matter component, with fraction $\alpha_{smooth}=1-\alpha_{star}-\alpha_{BH}$. The ($\kappa, \gamma$) values of each macro-image of the sample are taken from Schechter et al. (2014; see their Table 4).
    
    In order to minimize the sample variance, we compute a total of 100 magnification maps for each image using the Inverse Polygon Mapping algorithm (Mediavilla et al. 2006, 2011), resulting in a total of 3x7x7x40x100=588000 magnification maps. This procedure demanded a great deal of calculation time ($\sim$50000 CPU hours) requiring high-throughput computing services\footnote{PROTEUS Scientific Computing Cloud: \href{https://proteus.ugr.es}{https://proteus.ugr.es}; HTCondor: \href{https://htcondor.org}{https://htcondor.org}} for accelerating the map calculation process. The resolution of the maps was conservatively taken as 0.5 lt-day/pixel with a size for the maps of 250x250 pixels. Both, the pixel size and map resolution, were carefully selected and tested\footnote{We tested maps with sizes between 100 and 1500 pixels (aiming at the largest possible size) and found that 250 pixels provided the best balance between execution time and statistical completeness.} to provide a fair balance between the two mass components, so that we have a representative number of the heavy stars while keeping a manageable number of the lighter BHs for all the explored range of values of \{$\alpha_{star}, \alpha_{BH}$\}. A convolution with a Gaussian source of 1 lt-day (representative of the size of the X-ray source) is finally applied.
    The histograms of images $i$ and $j$ for each value of the parameters ($\alpha_{star}, \alpha_{BH}$) are used to calculate the probability density function (PDF) of observing a microlensing magnification $\Delta m_{ij}$, ${p_{klij}({\alpha_{BH}}_k,{\alpha_{star}}_l|\Delta m_{ij})}$ via cross-correlation (see Mediavilla et al. 2009).

     Finally, to obtain the corresponding PDFs for the abundance of BHs of planetary to substellar mass, we apply a Bayesian inference analysis, as explained in Esteban-Guti{\'e}rrez et al. (2022b). This method applies a statistical approach which calculates the posterior probability distribution of some parameters given a set of observed variables, which, in our case, are the microlensing magnifications. The global PDF is calculated as the product for image pairs,
    \begin{equation}
        p_{kl}({\alpha_{BH}}_k,{\alpha_{star}}_l)\propto \prod_{ij}{p_{klij}({\alpha_{BH}}_k,{\alpha_{star}}_l|\Delta m_{ij})}
    \end{equation} \\

\section{Results: Likelihood of the Bimodal Distribution of Abundances}

    The final PDFs and the marginalized PDFs for the surface mass density fraction of stars, $\alpha_{star}$, and BHs, $\alpha_{BH}$, are presented in Figure 1 for the three considered BH masses. The behaviour of the 2D PDF is as expected: we see a substantial impact of the BHs with lower masses, while their potential contribution decays as the BH mass increases to values closer to the stellar mass. For the highest considered mass of the BHs, they are not distinguishable from the stars and the joint PDF shows a strong degeneracy.  On the other extreme, the lowest mass BHs have more room to ``hide'', resembling the behaviour of smooth matter, and producing a much flatter distribution. The maximum probability for the fraction of stars is located at $\sim 0.1$ in all cases, with an expected value of $\sim 0.12$. A low contribution of the BHs peaking at (or near) $\alpha_{star}=0$ is obtained for all the masses considered with expected values of $\sim 0.1$ for $M_{BH}=0.0024M_{\odot}$ and $\lesssim 0.04$ for the remaining BH masses. For the BHs, we find upper limits of $\alpha_{BH} < [0.34,0.13,0.09]$ at the 68\% confidence level ($\alpha_{BH} < [0.64,0.27,0.16]$ at the 90\%) for $M_{BH}/M_{\odot}=0.0024, 0.013$ and $0.082$, respectively. 
    
    The main result of this paper is shown in Figure 2, where we show the results from the analysis of X-ray data in the present work together with the previous results based on optical observations presented in Esteban-Guti\'errez  et al. (2022b)\footnote{Notice that we have added a new point for a BH mass of $3M_{\odot}$ to the results obtained from the optical data, in order to fill the large mass gap between the largest mass considered in the present work of $0.08M_{\odot}$ and the lowest mass considered in Esteban-Gutiérrez et al. 2022b, of $10M_{\odot}$. This new point is also shown in Figure 2.}, to provide upper limits to the fraction of PBHs as dark matter, $f_{PBH} = \Omega_{PBH}/\Omega_{DM} = \alpha_{BH}/(\alpha_{BH}+\alpha_{smooth})$, for the mass range from $0.0024M_{\odot}$ up to $60M_{\odot}$ (with a new added point for a BH mass of $3M_{\odot}$). Figure 2 shows the ``confusion band'' for stellar mass BHs between the HBL and $3M_{\odot}$ for which microlensing cannot discriminate between stars and BHs.     
    \begin{figure}[H]
        \hspace{-0.1in}
        \includegraphics[width=1.05\linewidth]{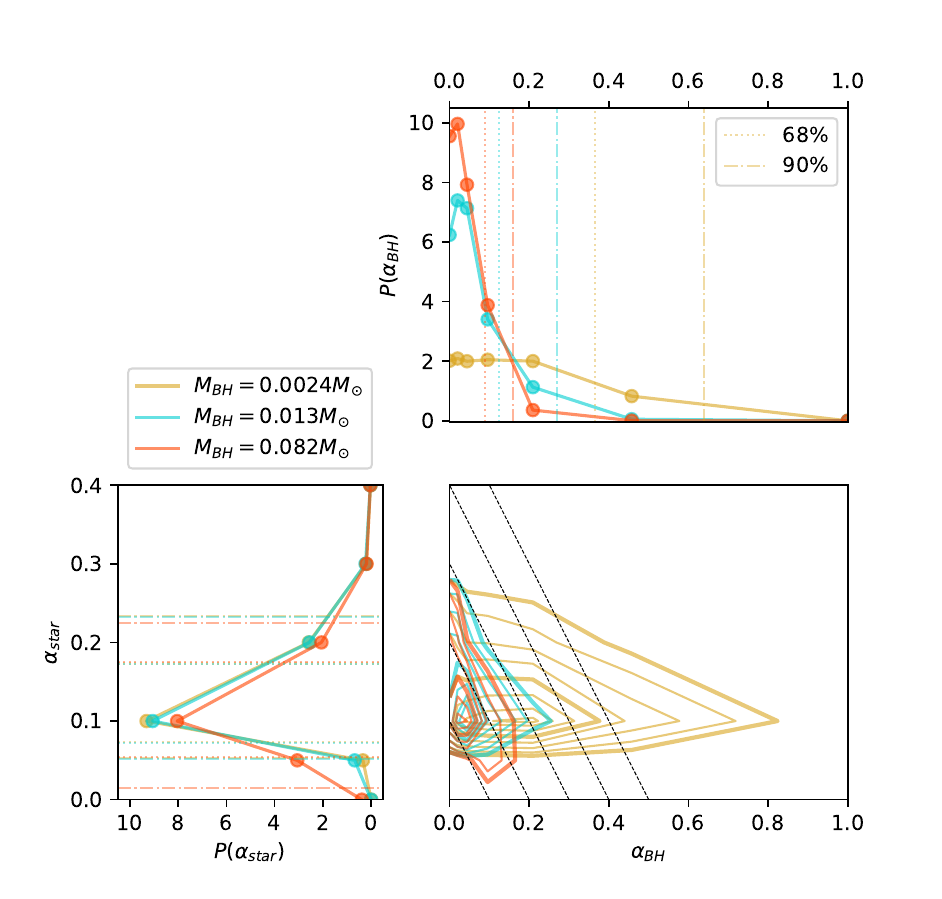}
        \caption{Probability distributions of the fraction of total mass density in stars and BHs of three masses: $M_{BH}/M_{\odot}=0.0024 ,0.013$ and $0.082$, plotted in yellow, turquoise and orange, respectively. Bottom right: Joint (2D) probability density function, $p_{ijkl}({\alpha_{BH}}_k,{\alpha_{star}}_l|\{\Delta m_{ij}\})$. Contour levels in steps of 0.25$\sigma$ with thicker lines for 1$\sigma$ and 2$\sigma$. Straight black dashed lines represent constant $\alpha_{smooth}$ for $\alpha_{smooth}=0.9,0.8,0.7,0.6,0.5$ (bottom to top). Top right and bottom left: Marginalized (1D) probability density functions, $p_{k}(\alpha_{BH_k}|\{\Delta m_{ij}\})$ and $p_{l}(\alpha_{star_l}|\{\Delta m_{ij}\})$,  of the fraction of total mass density in BHs, $\alpha_{BH}$, and stars, $\alpha_{star}$, respectively. Upper limits at the 68\% and 90\% confidence levels for each BH mass are indicated as dotted and dash-dotted lines, respectively.  \label{corner_plot_all}}
    \end{figure}

    \begin{figure}[H]
        \centering
        \includegraphics[width=0.85\linewidth]{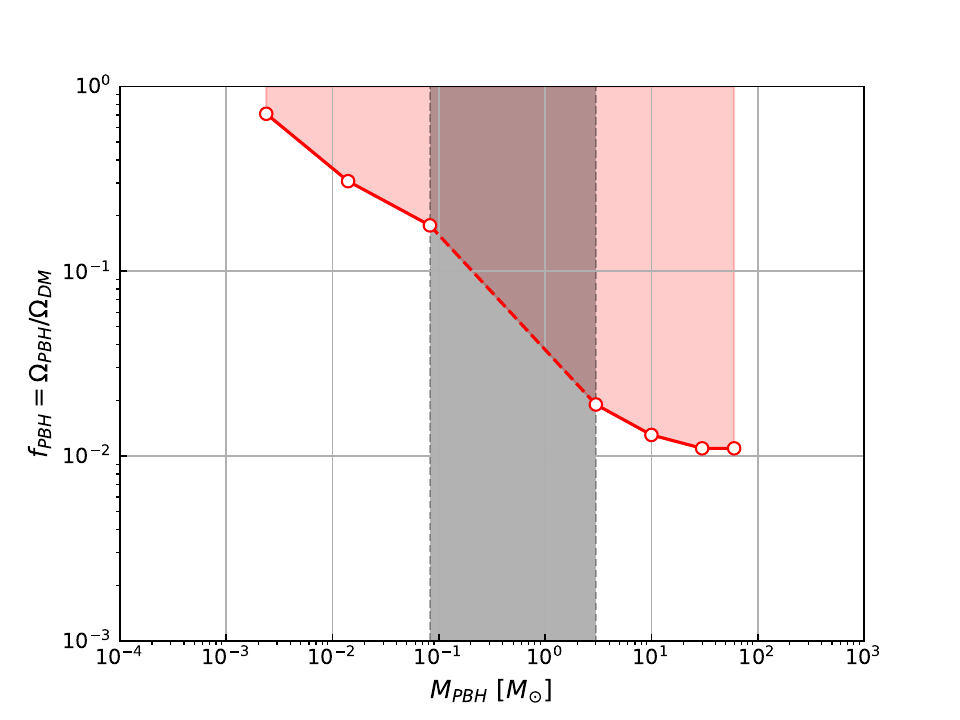}
        \caption{Upper limits of the contribution of PBHs to the fraction of dark matter from both, X-rays (this study) and the optical (Esteban-Guti\'errez  et al. 2022b), plus a new extra point at $3M_{\odot}$. The grey band corresponds to the mass range for which quasar microlensing cannot distinguish between stars and BHs.}
        \label{fig:PBHbounds}
    \end{figure} 
\section{Discussion: Constraints on the Dark Matter fraction}
    
       Our results show that the allowed abundance from quasar microlensing observations of BHs depends strongly on the mass of the microlenses, with a significant increase of the permitted abundance in compact objects for the lowest explored BH mass of $0.0024M_{\odot}$, corresponding to an Einstein radius comparable to a typical X-ray source ($\sim 1$ lt-day). This result is not unexpected, since when the ratio between the masses of the two microlens components (stars and BHs) is large enough, there will be a selective washing out of the small mass component if its Einstein radius is close to the source size considered. Below this limiting mass, X-ray microlensing becomes rather insensitive to the presence of BHs. On the other hand, it is evident that in the region of coincidence with typical stellar masses (from the HBL to $1-3M_{\odot}$) there is a degeneracy between both populations (stars and BHs) and we are not able to distinguish between them.
       Nevertheless, as we know from other grounds (Jiang \& Kochanek 2007 and references therein) that the mass fraction in stars is about $10\%$, we can assume that the fraction of BHs must be placed somewhere in this grey band (see Figure 2) most likely close to the red dashed line connecting the points at the HBL and $M_{BH}=3M_{\odot}$. 
    
    In order to provide a global comparison with other works regarding the estimation of the dark matter fraction, we use the free Python code PBHbounds\footnote{\href{https://github.com/bradkav/PBHbounds/}{https://github.com/bradkav/PBHbounds/}} (Kavanagh 2019) to add our new contributions in optical and X-rays (shown as the grey area in Figure 3) to this collective study\footnote{This corresponds to a selected mass range ($10^{-12}-10^{9}M_{\odot}$) from the complete sample of results given by the PBHbounds repository, compatible with constraints from various experiments involving microlensing (HSC, OGLE, EROS), GWs (LIGO-Virgo), pulsars (NANOGrav) and studies in galaxy dynamics.} (shown as light blue, red and orange areas in Figure 3). As we can see from this plot, our studies based on quasar microlensing establish the strongest bounds in the mass range between $10^{-1}$ and $10^{2}M_{\odot}$. We also added new constraints to the dark matter fraction in the substellar to planetary mass range ($10^{-3}-10^{-1}M_{\odot}$) using the present X-ray quasar microlensing results.\\

    \begin{figure}[H]
        \centering
        \includegraphics[width=\linewidth]{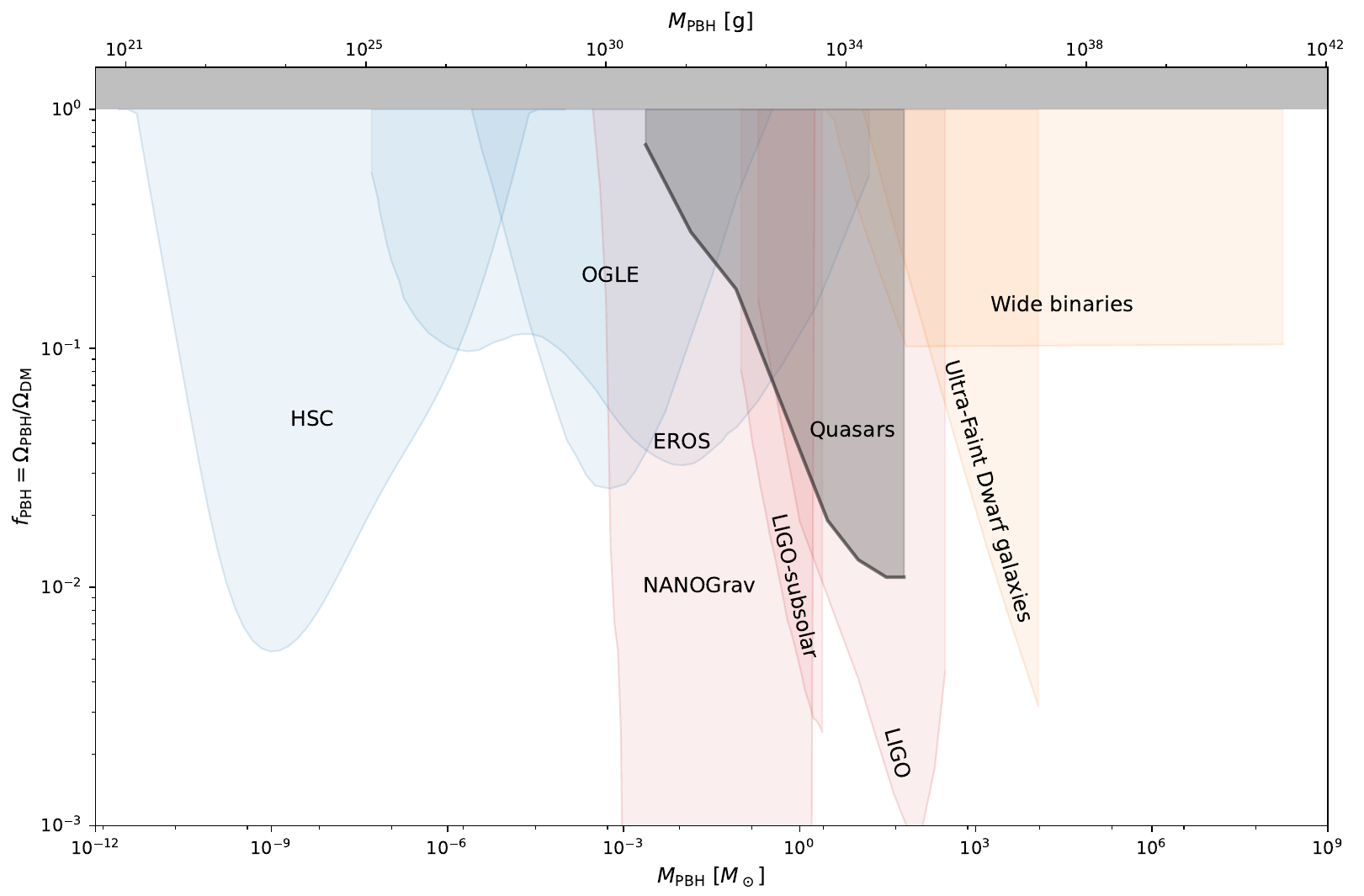}
         \caption{Dark matter fraction of PBHs inferred from microlensing (light blue) from OGLE, EROS and HSC collaborations, GWs/pulsars (red) from LIGO-Virgo and NANOGrav collaborations and dynamical constraints (orange) from Ultra-faint Dwarf Galaxies and Wide Binaries constraints on MACHOs
         in terms of their mass, assuming a monochromatic mass function (data taken from PBHbounds). The grey curve represents the quasar microlensing results for the visible and X-ray measurements, taking into account the upper limits at the 90$\%$ of confidence level.}
        \label{fig:PBHbounds}
    \end{figure} 
\clearpage
\section{Conclusions \label{conclusion}}
From X-ray microlensing data available in the literature on which we perform a Bayesian analysis of the impact of a bimodal population of stars and BHs with masses in the planetary to stellar range, we can summarize the following conclusions:

- Independently of the BH mass considered, the abundance of stars remains almost invariable, with an average value of $\alpha_{star}=0.12_{-0.05}^{+0.05}$ at the 68$\%$ of confidence level and $\alpha_{star}=0.12_{-0.07}^{+0.11}$ at the 90$\%$. This result confirms previous estimates based on optical data (e.g. Jim{\'e}nez-Vicente et al. 2015a).

- In the range of masses considered (from $0.0024M_{\odot}$ and $0.08M_{\odot}$), the expected abundance of BHs is very small ($<4\%$), reaching a $10\%$ abundance at the lowest mass. Below this mass, the X-ray microlensing becomes progressively insensitive to the presence of a population of compact objects.

- Using jointly the X-ray and optical microlensing data, we have been able to provide limits on the abundance of BHs (and of any other type of compact object) in the planetary to intermediate-mass range (see Figure 3). These limits are the strongest ones available to date in the 0.01$M_{\odot}$ to 60$M_{\odot}$ mass range and indicates that the BH abundance is negligible in the $\sim 0.005M_{\odot}$ to $\sim 100M_{\odot}$ mass range. 

We thank the anonymous referee for valuable comments that helped improving this paper. This research was supported by the Spanish projects PID2020-118687GB-C33, PID2020-118687GB-C32 and PID2020-118687GB-C31 financed by MCIN/AEI/10.13039/501100011033. J.J.V. is also supported by projects FQM-108, P20\_00334 and A-FQM-510-UGR20/FEDER financed by Junta de Andaluc\'\i a. J.A.M. is also supported by the Generalitat Valenciana with the project of excellence Prometeo/2020/085. AEG thanks the support from grant FPI-SO from the Spanish MINECO (research project SEV-2015-0548-17-4 and predoctoral contract BES-2017-082319) and acknowledges support from ANID Fondecyt Postdoctorado with grant number 3230554.

\end{document}